\documentclass[aps,twocolumn,showpacs,preprintnumbers,prb]{revtex4}
\usepackage{graphicx}
\usepackage{amssymb}

\newcommand{\R}{\mathbf{r}}

\newcommand{\UP}{n_{\uparrow}}
\newcommand{\DN}{n_{\downarrow}}
\newcommand{\TUP}{\tau_{\uparrow}}
\newcommand{\TDN}{\tau_{\downarrow}}

\newcommand{\be}{\begin{equation}}
\newcommand{\ee}{\end{equation}}
\newcommand{\bea}{\begin{eqnarray}}
\newcommand{\eea}{\end{eqnarray}}
\newcommand{\bean}{\begin{eqnarray*}}
\newcommand{\eean}{\end{eqnarray*}}

\begin{document}

\title{The many-body exchange-correlation hole at metal surfaces} 
\author{Lucian A. Constantin{$^1$} and J. M. Pitarke$^{2,3}$}
\affiliation{
$^1$Department of Physics and Quantum Theory Group, Tulane University, New Orleans, LA 70118\\
$^2$CIC nanoGUNE Consolider, Tolosa Hiribidea 76, E-20018 Donostia - San Sebastian, Basque
Country\\
$^3$Materia Kondentsatuaren Fisika Saila (UPV/EHU), DIPC, and
Centro F\'\i sica Materiales (CSIC-UPV/EHU),\\
644 Posta kutxatila, E-48080 Bilbo, Basque Country\\}

\date{\today}

\begin{abstract}
We present a detailed study of the coupling-constant-averaged exchange-correlation
hole density at a jellium surface, which we obtain in the random-phase approximation (RPA) 
of many-body theory. We report contour plots of the exchange-only and exchange-correlation
hole densities, the integration of the exchange-correlation hole density over the surface plane, the 
on-top correlation hole, and the energy density. We find that the on-top correlation hole is 
accurately described by local and 
semilocal density-functional approximations. We also find that for electrons that 
are localized far outside the surface the main part of the corresponding exchange-correlation hole is 
localized at the image plane.  
\end{abstract}


\maketitle

\section{Introduction}
\label{sec1}

The exchange-correlation (xc) energy of a many-electron system is the only 
density functional that has to be approximated in the Kohn-Sham (KS) formalism 
of density-functional theory (DFT).\cite{KS} It is formally defined by the following 
equation derived from the Hellmann-Feynman theorem:~\cite{a5} 
\begin{equation} 
E_{xc}[n]=\frac{1}{2}\int d\R\int d\R'\int^1_0 
d\lambda\,\frac{\rho^{\lambda}_{2}(\R',\R)}{|\R-\R'|}-U[n],
\label{e1}
\end{equation}
where $n(\R)$ is the density of a spin-unpolarized system of $N$ electrons, 
$U[n]=(1/2)\int d\R 
n(\R)n(\R')/|\R-\R'|$ is the Hartree energy, and $\rho^{\lambda}_{2}(\R',\R)$ is the 
reduced two-particle density matrix 
\begin{eqnarray}
\rho^{\lambda}_{2}(\R',\R)=N(N-1)\sum_{\sigma,\sigma',...,\sigma_{N}}\int
d\R_{3}...d\R_{N}\nonumber\\
\times |\Psi^{\lambda}(\R'\sigma',\R\sigma,\R_{3}\sigma_{3},...,\R_{N}\sigma_{N})|^{2}.
\label{e2}
\end{eqnarray}
Here, $\Psi^{\lambda}(\R_{1}\sigma_{1},...,\R_{N}\sigma_{N})$ is the 
antisymmetric wavefunction that yields the density $n(\R)$ and minimizes
the expectation value of  
$\hat{T}+\lambda\hat{V}_{ee}$, where $\hat{T}=-\sum^N_{i=1}\nabla^2_i/2$ and 
$\hat{V}_{ee}=\frac{1}{2}\sum_i\sum_{j\ne i}\frac{1}{|\R_i-\R_j|}$ 
are the kinetic energy and the electron-electron interaction operators.
%describes the ground state of the electron system 
%at coupling strength $\lambda$ and yields the density $n(\R)$. 
Eq.~(\ref{e2}) shows that
$\rho_{2}^{\lambda}(\R',\R)d\R'd\R$ is the joint probability of finding an electron 
of arbitrary spin in $d\R'$ at $\R'$ and an electron of arbitrary spin in $d\R$
at $\R$, assuming that the Coulomb interaction is $\lambda/|\R-\R'|$.
In the case of noninteracting KS electrons (i.e., $\lambda=0$),
$\rho_{2}^{\lambda=0}(\R',\R)$ is the \emph{exchange-only} reduced two-particle density matrix 
that is expressible in terms of KS orbitals.
(Unless otherwise stated, atomic units are used throughout,
i.e., $e^2=\hbar=m_e=1$.)

Hence, the xc energy can be expressed as the electrostatic interaction between individual 
electrons and the corresponding (and sorrounding) coupling-constant-averaged xc hole density
$\bar{n}_{xc}([n];\R,\R')$, as follows
\begin{equation}
E_{xc}[n]=\int d\R\; e_{xc}(\R)=\frac{1}{2}\int d\R\int d\R'\frac{n(\R) 
\bar{n}_{xc}([n];\R,\R')}{|\R-\R'|},
\label{e3}
\end{equation}
where [see Eqs.~(\ref{e1}) and (\ref{e3})]:
\begin{equation}
\bar{n}_{xc}([n];\R,\R')=\frac{1}{n(\R)}\int^1_0d\lambda\, 
\rho_{2}^{\lambda}(\R',\R)-n(\R'),
\label{e4}
\end{equation}
and $e_{xc}(\R)$ is the xc energy density.
The xc hole density $n_{xc}([n];\R,\R')$ is the result of three effects: self-interaction 
correction to the Hartree approximation, Pauli exclusion principle, and the electron 
correlation due to Coulomb repulsion between electrons.

The adiabatic-connection fluctuation-dissipation theorem provides an elegant path to the
{\it exact} coupling-constant-averaged xc hole density,\cite{a109,HG,GL,LP} which can be 
written as follows\cite{PE2}
\begin{eqnarray}
\bar{n}_{xc}([n];\R,\R')=\frac{1}{n(\R)}[-\frac{1}{\pi}\int^{\infty}_{0}d\omega\int^1_0d\lambda\,
\chi^{\lambda}(\R,\R';\omega)\nonumber\\-n(\R)\delta(\R-\R')],
\label{e5}
\end{eqnarray}
where $\chi^{\lambda}(\R,\R';\omega)$ is the density-response function of the 
interacting system at coupling strength $\lambda$ and satisfies, in the framework of 
time-dependent density-functional theory (TDDFT), the following exact Dyson-type equation\cite{GDP}
\begin{eqnarray}
&&\chi_{\lambda}({\bf r},{\bf r}';\omega)=
\chi_0({\bf r},{\bf r}';\omega)+\int d\R_1\,d\R_2\,
\chi_0({\bf r},{\bf r}_1;\omega)\cr\cr
&\times& \left\{{\lambda\over |\R_1-\R_2|}+
f_{xc,\lambda}[n](\R_1,\R_2;\omega)\right\}\,
\chi_{\lambda}({\bf r}_2,{\bf r}';\omega).
\label{e6}
\end{eqnarray}
Here, $\chi_0({\bf r},{\bf r}';\omega)$ is the density-response function of
non-interacting KS electrons (which is exactly known in terms of KS
orbitals~\cite{GK}) and
$f_{xc,\lambda}[n]({\bf r},{\bf r}';\omega)$ 
is the Fourier transform with respect to time [$f_{xc,\lambda}[n]({\bf r},{\bf 
r}';\omega)=\int^{\infty}_{-\infty}dt e^{i\omega t}f_{xc,\lambda}[n]({\bf r},t,{\bf 
r}',0)$] of the {\it unknown}
$\lambda$-dependent xc kernel, formally defined by
\begin{equation}
f_{xc,\lambda}[n](\R,t,\R',t')=\frac{\delta v^{\lambda}_{xc}[n](\R,t)}{\delta
n(\R',t')},
\label{e7}
\end{equation}
%
%\begin{equation}
%f^{\lambda}_{xc}[n](\R,\R';\omega)=\frac{\delta v^{\lambda}_{xc}[n](\R',\omega)}{\delta 
%n(\R,\omega)},
%\label{e7}
%\end{equation}
where $v^{\lambda}_{xc}[n](\R,t)$ is the exact time-dependent xc potential of TDDFT. When
$f_{xc,\lambda}[n]({\bf r},{\bf r}';\omega)$ is taken to be zero,
Eq.~(\ref{e6}) reduces to the random phase approximation (RPA).
If the interacting  density response function
$\chi_{\lambda}({\bf r},{\bf r}';\omega)$ is replaced
by the noninteracting KS density-response function
$\chi_0({\bf r},{\bf r}';\omega)$, then Eq. (\ref{e5})
yields the {\it exchange-only} hole density.

The scaling relation of the correlation hole density at coupling constant $\lambda$ 
\cite{a41,a48} leads to the following equation for the coupling-constant-averaged 
correlation hole density:
\begin{equation}
\bar{n}_{c}([n];\R,\R')=\int^1_0 d\lambda\, (\frac{\lambda}{w})^3 
n^w_c([n_{w/\lambda}],\frac{\lambda}{w}\R,\frac{\lambda}{w}\R'),
\label{e8}
\end{equation}
where $0<w<<1$ is a fixed constant, and $n_\gamma(\R)=\gamma^3n(\gamma\R)$ is a
uniformly-scaled density.\cite{a42} Eq. (\ref{e8}) shows that the whole many-body problem is 
equivalent 
to the knowledge of the universal correlation hole density at a small, fixed coupling 
strength $w$.   

There is a "Jacob's ladder" \cite{jacob} classification (in RPA and beyond RPA) of 
nonempirical approximations to the angle-averaged xc hole 
density
\begin{equation}
\bar n_{xc}([n];{\bf r},u)=
{1\over 4\pi}\int d\Omega\,\bar n_{xc}([n];{\bf r},{\bf r}'),
\label{e9}
\end{equation}
where $d\Omega$ is the differential solid angle around the direction of
${\bf u}={\bf r}'-{\bf r}$. The simplest rung of the ladder is the local spin density 
approximation (LSDA) of the xc hole density $\bar n_{xc}(\UP,\DN;u)$ that has as ingredients 
only the spin densities.
(For the RPA-based LSDA xc hole and for the LSDA xc hole, see Refs.~\onlinecite{EP,YPK} and 
Refs.~\onlinecite{PW,EP,a112}, respectively.) 
The next rung is the generalized gradient approximation (GGA) xc hole density
$\bar n_{xc}(\UP,\DN,\nabla\UP,\nabla\DN,u)$. (See Ref.~\onlinecite{EP} for the smoothed GGA 
exchange hole model, Ref.~\onlinecite{PBW} for the 
PBE-GGA\cite{PBE} correlation hole, and Ref.~\onlinecite{YPK} for the RPA-based GGA hole 
model. For a GGA xc hole constructed for solids, see Ref.~\onlinecite{CPP}.) The third rung on this 
ladder is the non-empirical meta-GGA xc hole density\cite{CPT}
$\bar n_{xc}(\UP,\DN,\nabla\UP,\nabla\DN,\TUP,\TDN,u)$ that depends on spin densities and 
their gradients, as well as the positive KS kinetic energy densities $\TUP$ and 
$\TDN$, and that was constructed to satisfy many exact constraints. 
(For an RPA-based meta-GGA xc hole model, see also Ref.~\onlinecite{CPT}.)

Jellium is a simple model of a simple metal, in which the ion cores are replaced by a 
uniform positive background of density
$\bar{n}=3/4\pi r^{3}_{s}=k^{3}_{F}/3\pi^{2}$ and the valence electrons in the 
spin-unpolarized bulk neutralize this background. $r_{s}$ is the bulk density
parameter and $k_{F}$ is the magnitude of the bulk Fermi wavevector. At a jellium 
surface, the plane $z=0$
separates the uniform positive
 background ($z>0$) from the vacuum ($z<0$), and the electrons can leak out into 
the vacuum. This
electron system is translationally invariant in the plane of the surface.

The exchange hole at a jellium surface was studied in
Ref.~\onlinecite{SB} (using a finite linear-potential model\cite{SMF}),
and in Refs.~\onlinecite{Ju,MM} (using the infinite barrier model (IBM)~\cite{Ne}). The 
behavior of the xc hole at a jellium surface was investigated at the RPA level using 
IBM orbitals.\cite{IM} Hence, existing calculations of the exchange-only and xc hole at a 
jellium surface invoke either a finite linear-potential model or the IBM for the description 
of single-particle orbitals. An exception is a self-consistent calculation of the RPA xc hole 
density reported briefly in Refs.~\onlinecite{PE1} and \onlinecite{NP}, in which accurate LSDA 
single-particle orbitals were employed.

In this paper, we present extensive self-consistent calculations of the {\it exact}-exchange 
hole and the RPA xc hole at a jellium surface. We report contour plots of the corresponding 
hole densities, the integration of the xc hole density over the surface plane, and the on-top 
correlation hole. We find that the on-top RPA correlation hole $\bar{n}_c([n];\R,\R)$ is 
accurately described by the on-top RPA-based LSDA hole, in accord with the work of
Perdew \emph{et al}.\cite{LP,BPL,BPE} 

\section{The exact-exchange hole and the RPA xc hole at a jellium surface}
\label{sec2}

Let us consider a jellium surface with the surface plane at $z=0$. Using its 
translational invariance in a plane perpendicular to the $z$ axis, the 
coupling-constant-averaged xc hole density of Eq.~(\ref{e5}) can be written as follows\cite{NP}
\begin{eqnarray}
&\bar n_{xc}([n];r,z,z')=-\frac{1}{2\pi}\int^\infty_0 d q_{||} \;q_{||} 
J_0(q_{||}r)[\frac{1}{\pi n(z)}\int^1_0d\lambda\int^\infty_0d\omega\nonumber\\
&\times\chi^{\lambda}(q_{||},z,z',\omega)-\delta(z-z')],
\label{e11}
\end{eqnarray}
where $r=|\R_{||}-\R_{||}'|$, and $\bf{q}_{||}$ is a two-dimensional (2D) wavevector.
$\chi^{\lambda}(q_{||},z,z',\omega)$ represents the 2D Fourier transform of the interacting 
density response function of Eq.~(\ref{e6}), which in the RPA is obtained by neglecting the xc 
kernel $f_{xc}$. The exact-exchange hole density is obtained by simply replacing in 
Eq.~(\ref{e11}) $\chi^{\lambda}(q_{||},z,z',\omega)$ by the corresponding KS noninteracting 
density response function $\chi^{0}(q_{||},z,z',\omega)$. 

For the evaluation of Eq.~(\ref{e11}), we follow the method described in 
Ref.~\onlinecite{PE2}. We consider a jellium slab, and we assume that the electron density 
$n(z)$ vanishes at a distance $z_0=2\lambda_F$ ($\lambda_F=2\pi/k_F$ is the  
bulk Fermi wavelength) from either jellium edge.\cite{note1} We expand the single-particle 
wave functions entering the evaluation of $\chi^{0}(q_{||},z,z',\omega)$ in a sine Fourier 
representation, and the density-response functions $\chi^{0}(q_{||},z,z',\omega)$ and
$\chi_\lambda(z,z';q_\parallel,\omega)$ in a double-cosine Fourier representation.  We also 
expand the Dirac delta function entering Eq.~(\ref{e11}) in a double-cosine representation 
(see Eq. (A2) of Ref.~\onlinecite{PE2}). We take all the
occupied and unoccupied single-particle orbitals and energies to be the LSDA eigenfunctions 
and eigenvalues of a KS Hamiltonian, as obtained by using the Perdew-Wang parametrization 
\cite{PW1} of the Ceperley-Alder xc energy of the uniform electron gas.\cite{ca}

In the calculations presented below, we have considered jellium slabs with several bulk 
parameters $r_s$ and a thickness $a=2.23\,\lambda_F$ for the positive background.
For $r_s=2.07$, such slab corresponds to about four atomic layers of Al(100) and it was used 
in the wavevector analysis of the RPA \cite{PCP} and beyond-RPA\cite{CPP,CPDGLP} xc surface energy.

%
%%%%%%%%%%%%%%%%%%%%%%%%%%%%%%%%%%%%%%%%%%%%%%%%%%%%%
\begin{figure}
\includegraphics[width=\columnwidth]{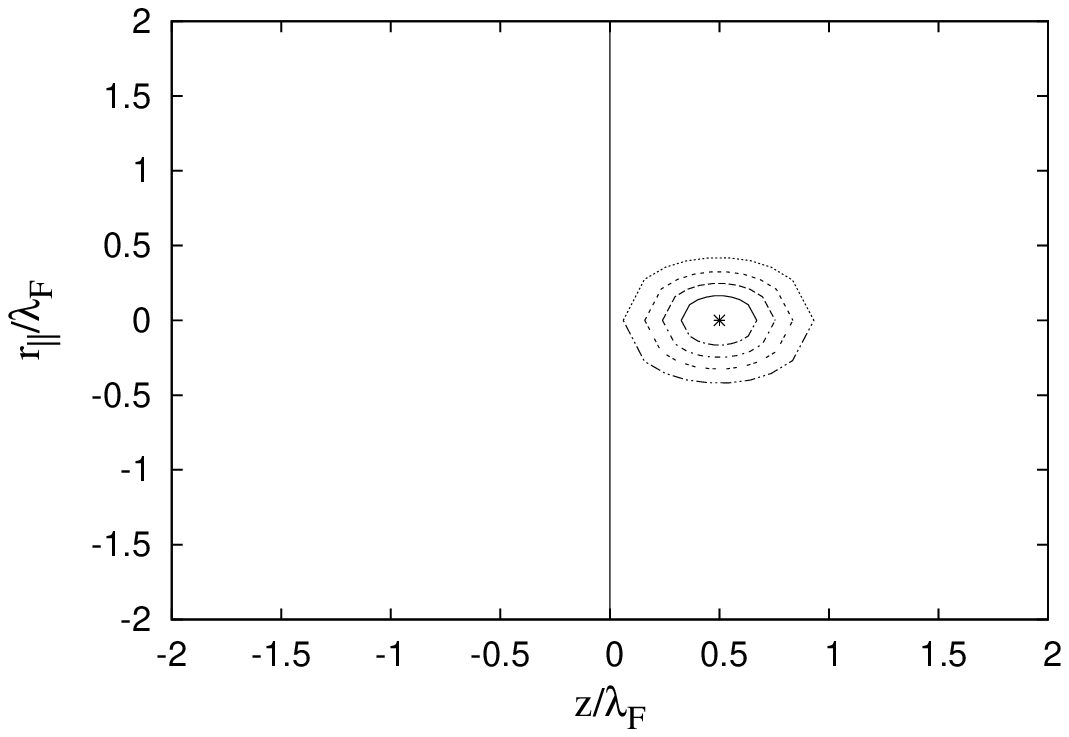}
\includegraphics[width=\columnwidth]{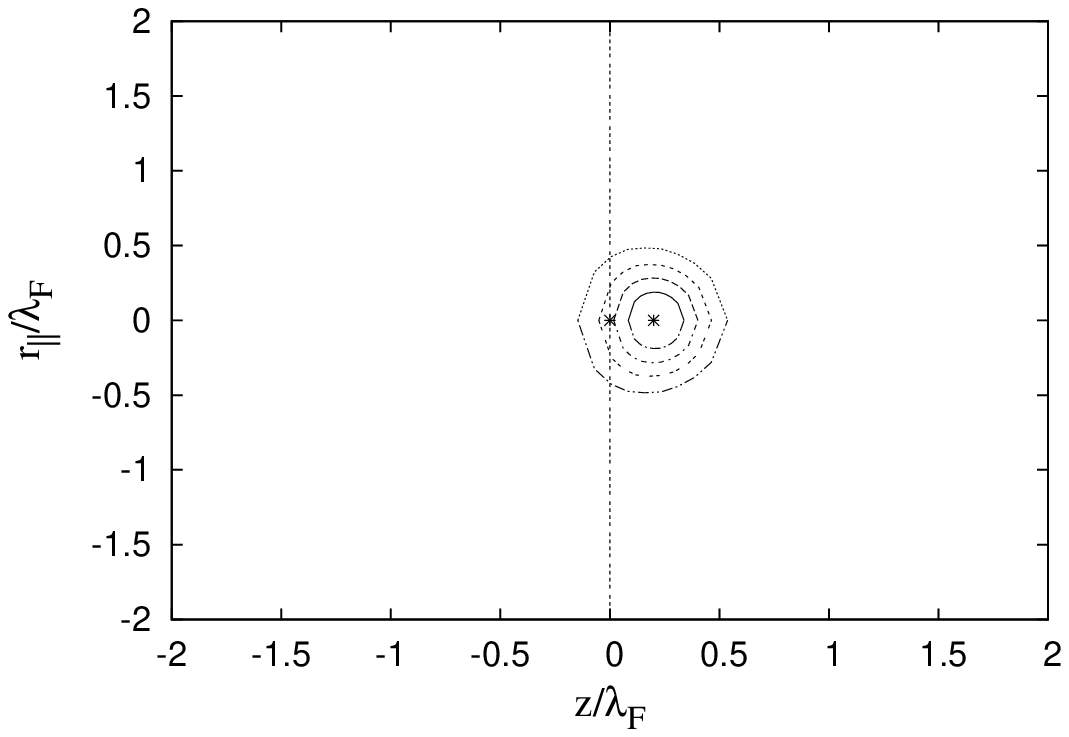}
\includegraphics[width=\columnwidth]{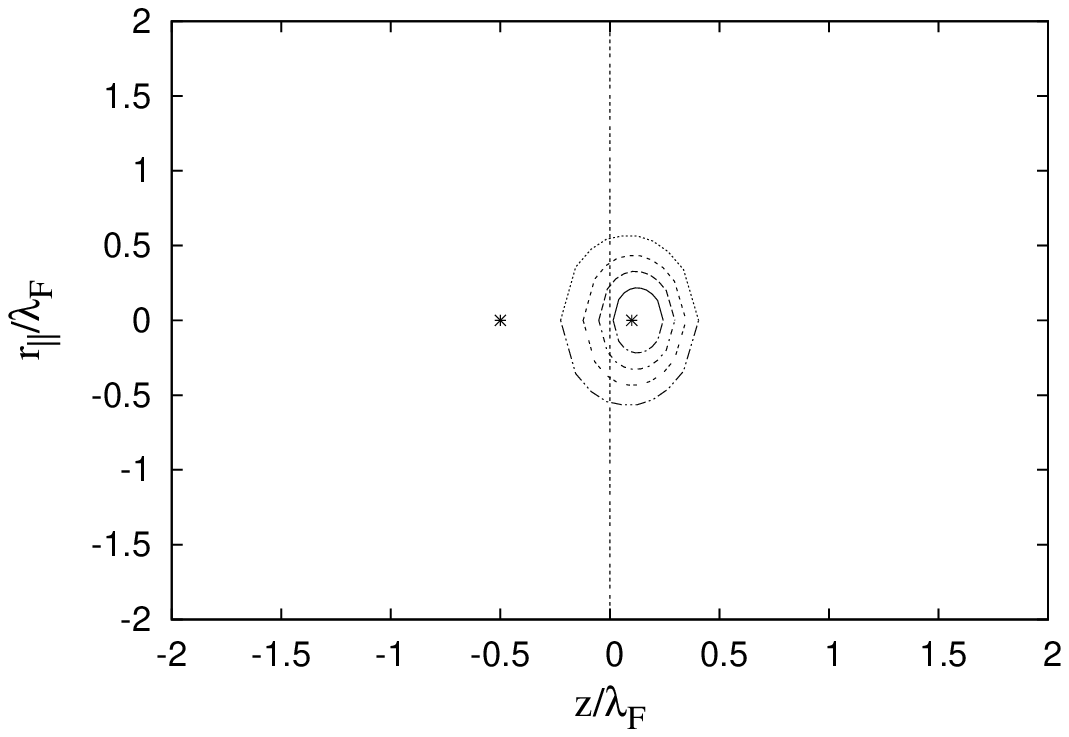}
\includegraphics[width=\columnwidth]{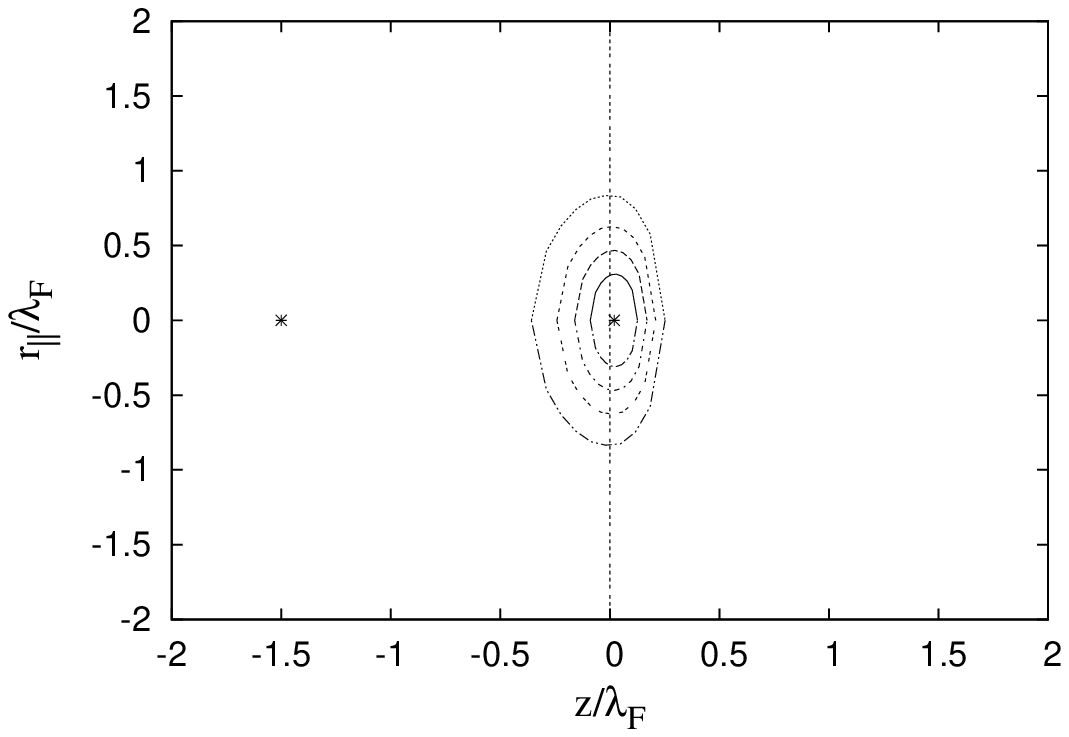}
\caption{ Contour plots of the exchange hole density $\bar{n}_x(r_{||},z,z')$ for several fixed 
values of the electron position: $z = 0.5\lambda_F$ (inside the bulk), $z = 0$ (on 
the surface), $z = -0.5\lambda_F$ (in the vacuum) and $z = -1.5\lambda_F$ (far outside the 
surface in the vacuum). The bulk parameter is $r_s=2.07$, the jellium surface is at 
$z=0$, and $r_{||}=\pm|\R_{||}-\R_{||}'|$.}
\label{f1}
\end{figure}
%%%%%%%%%%%%%%%%%%%%%%%%%%%%%%%%%%%%%%%%%%%%%%%%%%%%%%%
%
%
%%%%%%%%%%%%%%%%%%%%%%%%%%%%%%%%%%%%%%%%%%%%%%%%%%%%%
\begin{figure}
\includegraphics[width=\columnwidth]{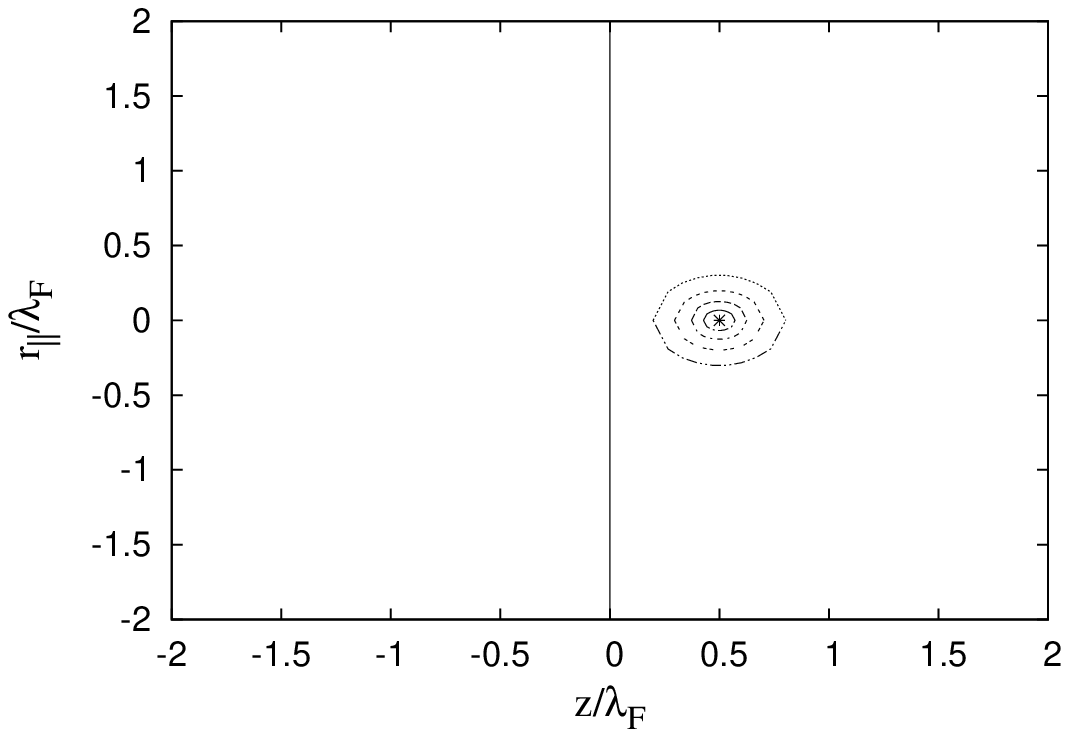}
\includegraphics[width=\columnwidth]{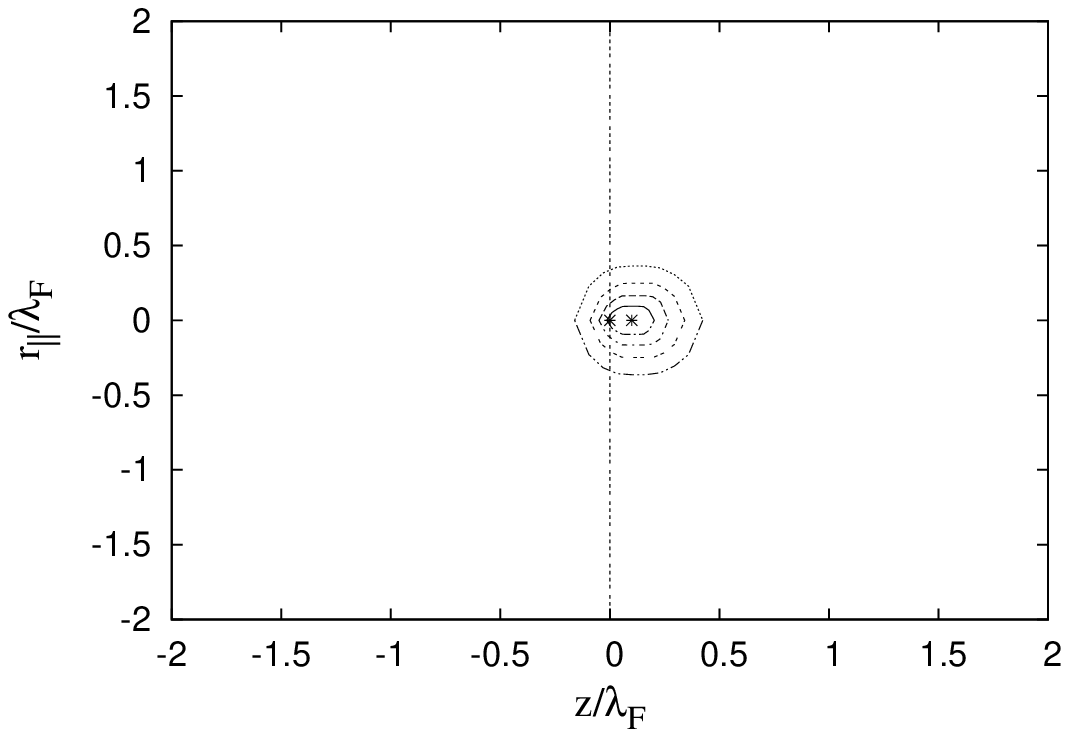}
\includegraphics[width=\columnwidth]{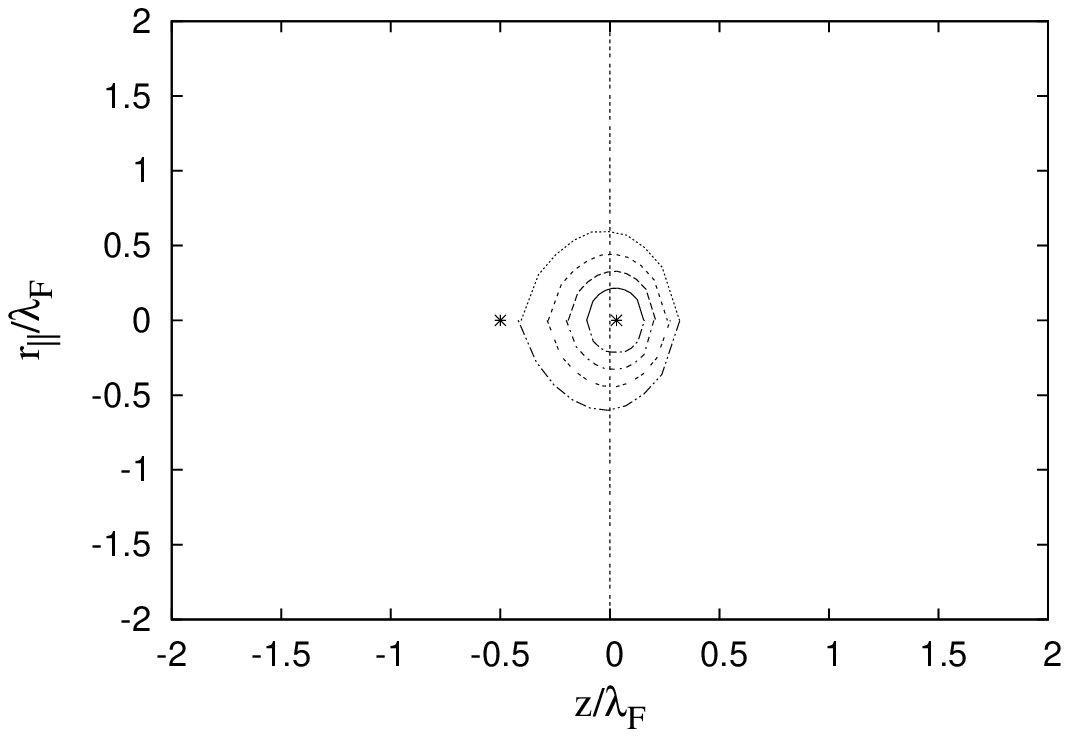}
\includegraphics[width=\columnwidth]{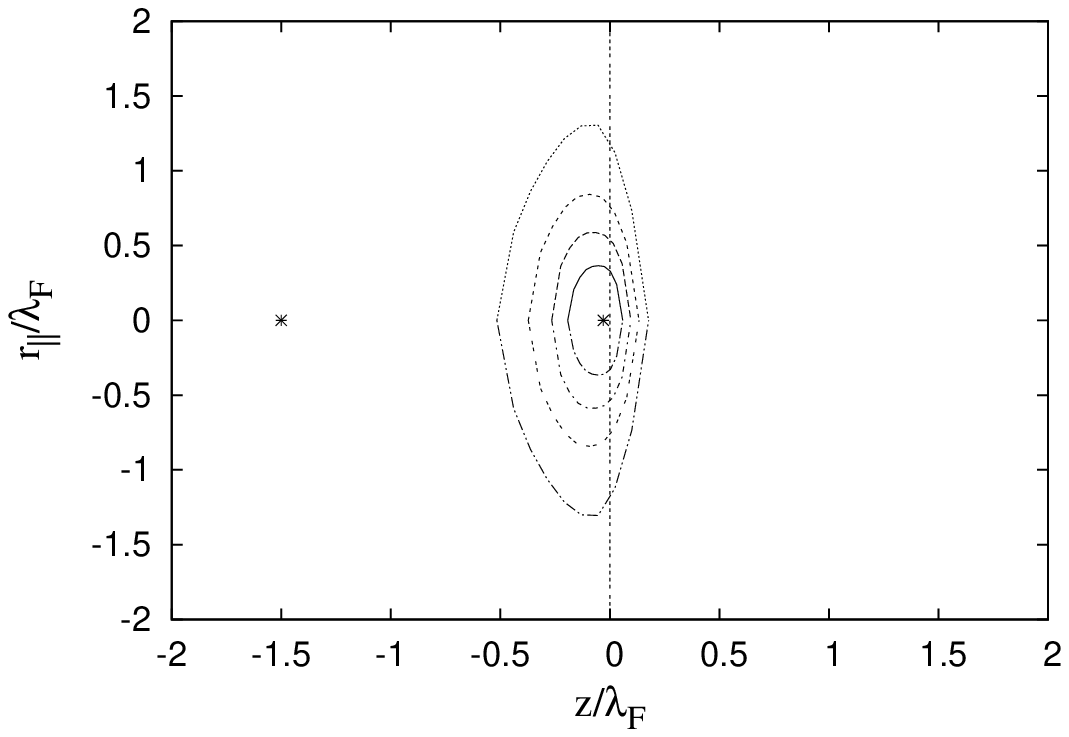}
\caption{ Contour plots of the RPA coupling-constant-averaged xc hole density 
$\bar{n}_x(r_{||},z,z')$ for several 
fixed
values of the electron position: $z = 0.5\lambda_F$ (inside the bulk), $z = 0$ (on
the surface), $z = -0.5\lambda_F$ (in the vacuum) and $z = -1.5\lambda_F$ (far outside the 
surface in the vacuum). 
The bulk parameter is $r_s=2.07$, the jellium surface is at
$z=0$, and $r_{||}=\pm|\R_{||}-\R_{||}'|$.
See also Fig. 1 of Ref. \cite{NP}.}
\label{f2}
\end{figure}
%%%%%%%%%%%%%%%%%%%%%%%%%%%%%%%%%%%%%%%%%%%%%%%%%%%%%%%
%

In Figs.~\ref{f1} and \ref{f2}, we show contour plots for the exact-exchange hole density and 
the self-consistent RPA xc hole density, respectively. In the bulk, both the exchange-only 
hole and the xc hole are spherical and the xc hole is more localized, as in the case of a 
uniform electron gas. Near the surface, both the exchange-only hole and the xc hole happen to 
be distorted, the center of gravity being closer to the surface when correlation is included. 
For an electron that is localized far outside the surface, the corresponding exchange-only 
hole and xc hole remain localized near the surface; Figs.~\ref{f1} and \ref{f2} show that the 
introduction of correlation results in a flatter hole, which in the case of an electron that 
is infinitely far from the surface becomes completely localized at a plane parallel to the 
surface. This is the image plane. We recall that the RPA xc hole density is exact in the limit 
of large separations (where $u=|\R-\R'|\to\infty$), and yields therefore the exact location of 
the image plane. 

%
%%%%%%%%%%%%%%%%%%%%%%%%%%%%%%%%%%%%%%%%%%%%%%%%%%%%%
\begin{figure}
\includegraphics[width=\columnwidth]{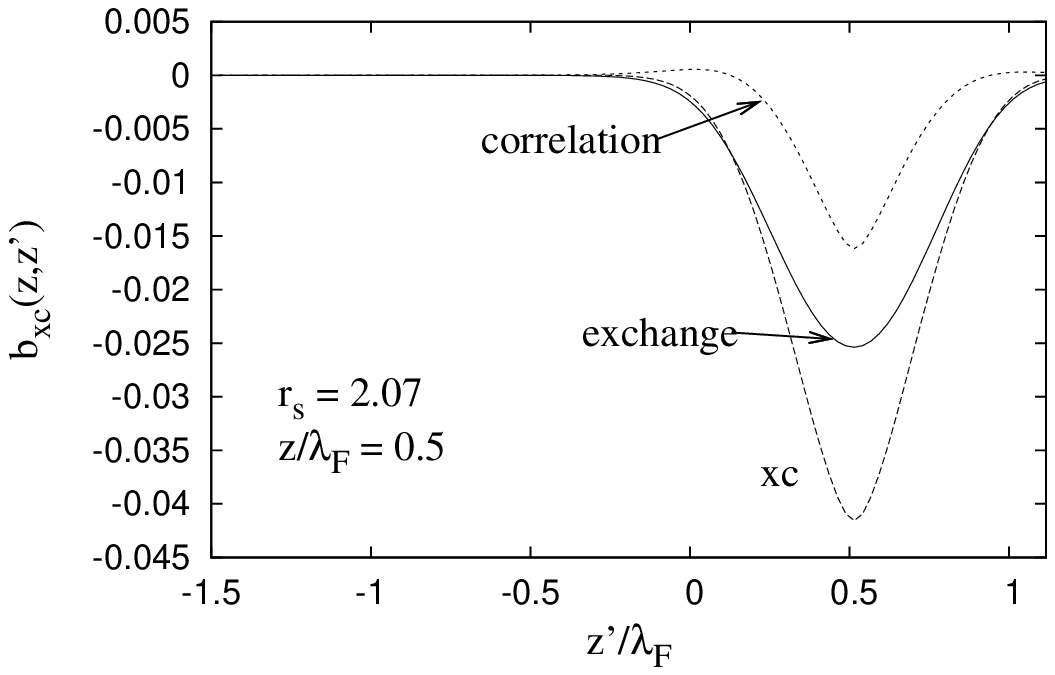}
\includegraphics[width=\columnwidth]{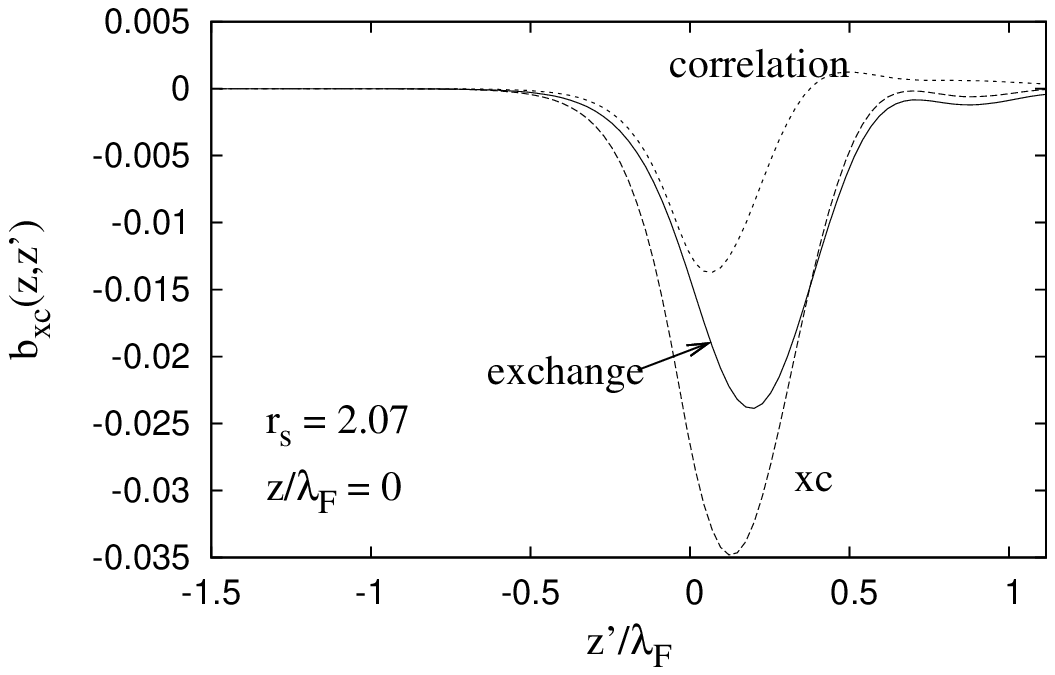}
\includegraphics[width=\columnwidth]{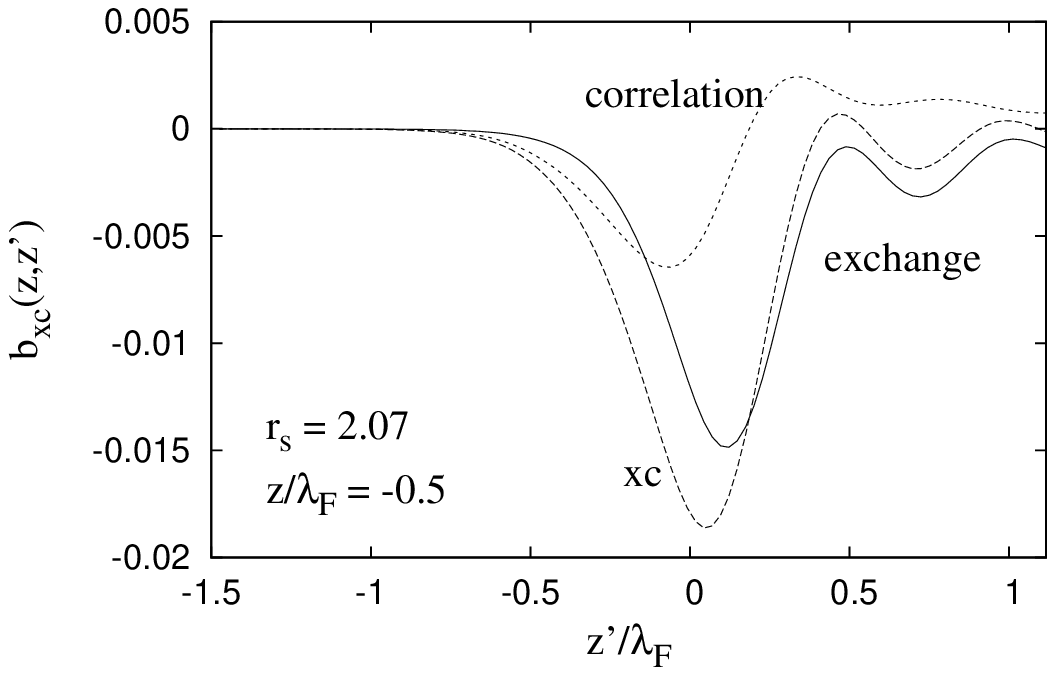}
\includegraphics[width=\columnwidth]{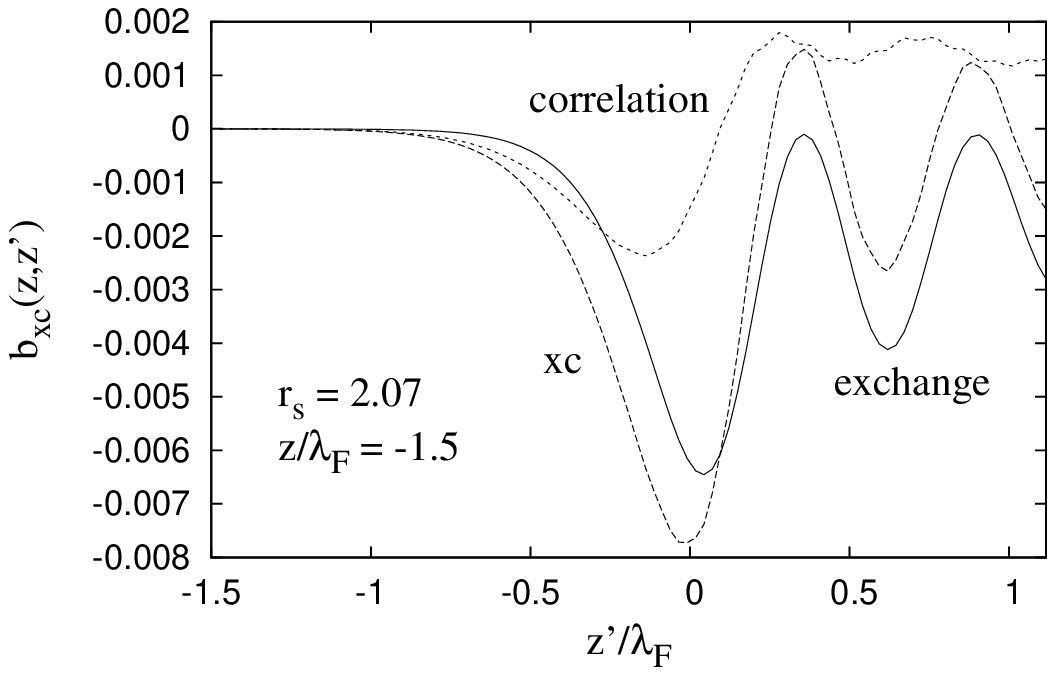}
\caption{ 
$b_{xc}(z,z')$ of Eq. (\ref{e12}) versus $z'/\lambda_F$ for the same positions of the 
electron as in Figs. \ref{f1} and  \ref{f2}.
The bulk parameter is $r_s=2.07$ and the jellium surface is at
$z=0$.}
\label{f3}
\end{figure}
%%%%%%%%%%%%%%%%%%%%%%%%%%%%%%%%%%%%%%%%%%%%%%%%%%%%%%%
%

The integration of the xc hole density over the whole surface plane, 
\begin{equation}
b_{xc}([n],z,z')=\int^\infty_0 dr\; \bar{n}_{xc}([n];r,z,z'),
\label{e12}
\end{equation}
represents a quantity of interest for a variety of theoretical and experimental situations 
(see for example Refs.~\onlinecite{NFN,CC11}). Below we show that $b_{xc}([n];z,z')$ represents a 
suitable quantity to describe the behavior of the xc hole corresponding to a given electron 
located at an arbitrary distance from the surface. In Fig.~\ref{f3}, we plot this quantity, 
versus $z'$, for $r_s=2.07$ and a given electron located at $z=0.5\lambda_F$, $z=0$, 
$z=-0.5\lambda_F$, and $z=-1.5\lambda_F$. We see from this figure that (i) correlation 
damps out the oscillations that the exchange hole exhibits in the bulk part of the surface, 
and (ii) in the case of a given electron located far from the surface into the vacuum the main 
part of the exchange-only and the xc hole is found to be near the surface (see also 
Figs.~\ref{f1} and \ref{f2}), although the exchange-only hole appears to be much more 
delocalized with a considerable weight within the bulk. 

%
%%%%%%%%%%%%%%%%%%%%%%%%%%%%%%%%%%%%%%%%%%%%%%%%%%%%%
\begin{figure}
\includegraphics[width=\columnwidth]{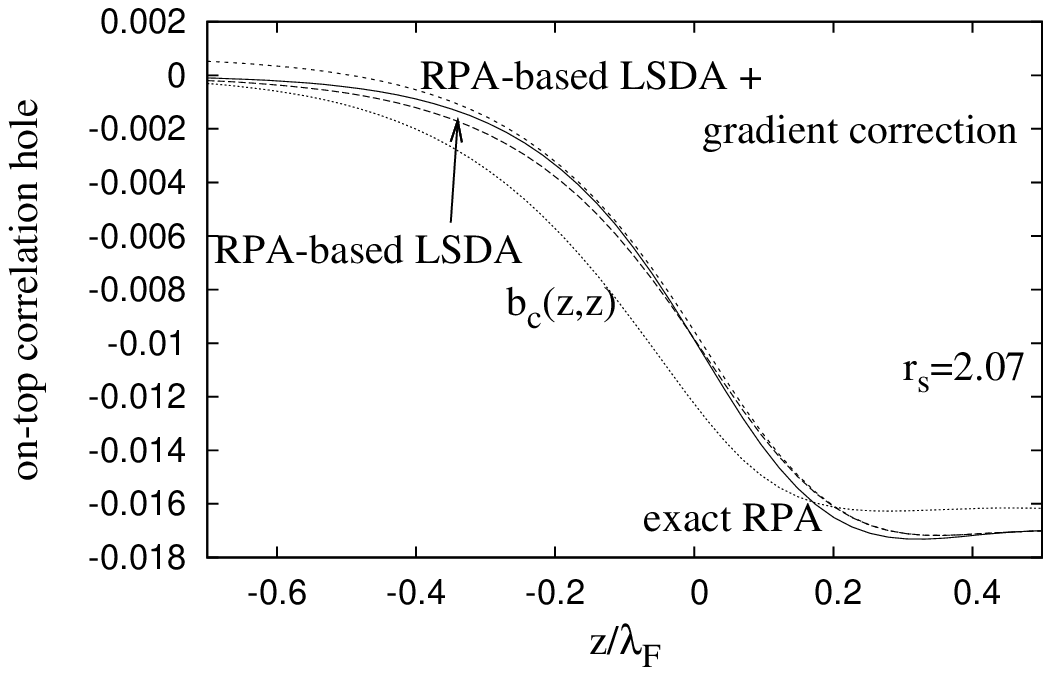}
\caption{ 
On-top coupling-constant-averaged correlation hole $\bar{n}_c(\R,\R)$ at a jellium 
surface. Also shown is $b_c(z,z)$ of Eq. (\ref{e12}). The bulk parameter is $r_s=2.07$ and 
the jellium surface is at $z=0$. }
\label{f4}
\end{figure}
%%%%%%%%%%%%%%%%%%%%%%%%%%%%%%%%%%%%%%%%%%%%%%%%%%%%%%%
%

Let us now focus on the on-top xc hole. The LSDA accurately accounts for short wavelength 
contributions to the xc energy; \cite{BPL} thus, all the nonempirical approximations of the xc hole 
have been constructed to recover the LSDA on-top xc hole $\bar{n}^{LSDA}_{xc}(\R,\R)$.
The slowly-varying electron gas was treated within RPA by Langreth and Perdew 
\cite{LP}. For a spin-unpolarized system, the gradient correction to the LSDA on-top 
correlaton hole density is \cite{BPE}
\begin{equation}
\bar{n}_c^{GEA}(\R,\R)=\bar{n}_c^{LSDA}(\R,\R)+\frac{|\nabla n|^2}{72\pi^3 n^2}.
\label{e13}
\end{equation}
In Fig. \ref{f4}, we show the on-top correlation hole for the exact RPA, the RPA-based LSDA
(see Ref.~\onlinecite{YPK}) and the RPA-based GEA of Eq.~(\ref{e13}). We see that for a jellium 
surface the RPA-based LSDA on-top correlation hole nearly coincides with the corresponding
exact RPA on-top correlation hole; this is in contrast with the case of strong inhomogeneous systems 
(e.g., Hooke's atom).\cite{BPL} The gradient correction of Eq.~(\ref{e13}) improves the already 
accurate RPA-based LSDA on-top correlation hole in the slowly-varying density region, but is 
inacurate in the tail of the density. Fig.~\ref{f4} also shows that the integrated
$b_c(z,z)$ of Eq.~(\ref{e12}) is more (less) negative in the vacuum (bulk) than the actual
on-top correlation hole.

At this point, we would like to emphasize that while the RPA on-top correlation hole in the bulk
is too negative but finite, the on-top correlation hole diverges in the bulk within a TDDFT scheme 
that uses a wavevector and frequency independet xc kernel like in the adiabatic local-density 
approximation (ALDA) 
\begin{equation}
f^{ALDA}_{xc,\lambda}[n](\R,\R',\omega)=\frac{d v^{\lambda,unif}_{xc}[n(\R)]}{d
n(\R)}\delta(\R-\R'),
\label{e10}
\end{equation}
or the energy-optimized local-density approximation of
Ref.~\onlinecite{DW}. (See the discussion after Eq.~(3.9) of Ref.~\onlinecite{DW}). 
Here, $v^{\lambda,unif}_{xc}[n(\R)]$ is the xc potential of a uniform electron gas
of density $n(\R)$. An xc kernel borrowed from a uniform-gas xc kernel that has the correct 
large-wavevector behavior (see, e.g., the xc kernels of Refs.~\onlinecite{CSOP,CP,PP}) would yield a
finite on-top correlation hole.   
%
%%%%%%%%%%%%%%%%%%%%%%%%%%%%%%%%%%%%%%%%%%%%%%%%%%%%%
\begin{figure}
\includegraphics[width=\columnwidth]{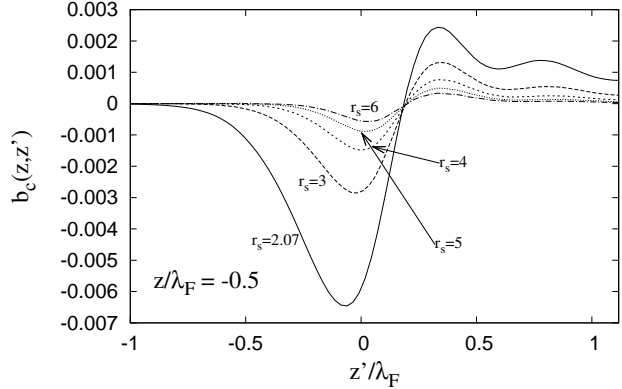}
\caption{
The correlation hole $b_c(z,z')$ of an electron at position $z=-0.5\lambda_F$ for several
values of the bulk parameter $r_s=2.07, 3, 4, 5,$ and $6$.
The jellium surface is at $z=0$.}
\label{f5}
\end{figure}
%%%%%%%%%%%%%%%%%%%%%%%%%%%%%%%%%%%%%%%%%%%%%%%%%%%%%%%
%
Fig. \ref{f5} shows the integrated correlation hole of Eq.~(\ref{e12}) for an electron at the vacuum 
side of the surface, at the position $z=-0.5\lambda_F$ and for several values of the electron-density 
parameter $r_s:$ 1.5, 2.07, 3, 4, 5, and 6. In the bulk, the correlation hole exhibits damped 
oscillations with $r_s$-dependent amplitude and a period that does not depend on the electron density 
and is close to the period ($\sim 0.56\lambda_F$) of the corresponding oscillations exhibited by the 
exchange-only hole. 

Finally, we look at the xc energy density $e_{xc}$ defined in Eq.~ (\ref{e3}). We note that adding to 
the actual $e_{xc}$ of Eq.~ (\ref{e3}) an arbitrary function of the position ${\bf r}$ that 
integrates to zero yields the same total xc energy.\cite{TSSP11} The Laplacian of the density 
$\nabla^2 n$ integrates to zero for finite systems, it plays an important role in the gradient 
expansion of the kinetic-energy density,\cite{CR,PC,Ki} and it is an important ingredient in the 
 construction of density-functional approximations for the kinetic energy density\cite{PC,CR} and the 
xc energy.\cite{PC}

We define the simplest possible Laplacian-level RPA-based LSDA (the RPA-based L-LSDA) xc energy density: 
\begin{equation}
e^{L-LSDA-RPA}_{xc}(\R)=e^{LSDA-RPA}_{xc}(\R)-C\nabla^2 n(\R),
\label{e14}
\end{equation}
where $C$ is a constant parameter which we find by minimizing the difference between 
$e^{RPA-L-LSDA}_{xc}$ and $e^{RPA}_{xc}$. We find  $C=0.3$ for a jellium slab with $r_s=2.07$, and 
its value gets larger as $r_s$ increases.
%
%%%%%%%%%%%%%%%%%%%%%%%%%%%%%%%%%%%%%%%%%%%%%%%%%%%%%
\begin{figure}
\includegraphics[width=\columnwidth]{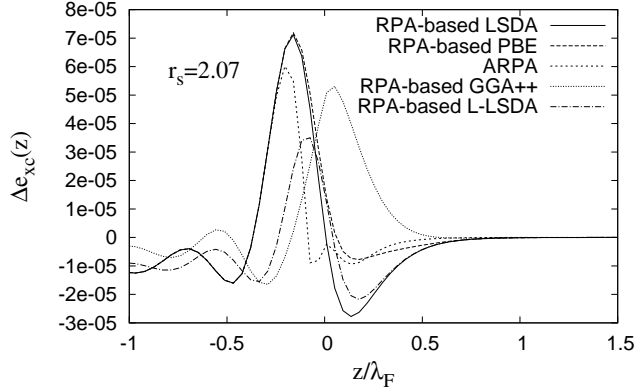}
\caption{
$\Delta e_{xc}(z)=e^{RPA}_{xc}(z)-e^{approx}_{xc}(z)$ versus $z/\lambda_F$ at a surface of a jellium 
slab, 
for several xc approximations: 
 RPA-based LSDA \cite{YPK}, RPA-based PBE \cite{YPK}, ARPA GGA \cite{CRP11}, RPA-based GGA++ 
\cite{CC11}, and RPA-based L-LSDA (Eq. (\ref{e14}) with $C=0.3$).
The bulk parameter is $r_s=2.07$, and the edge of the positive background is at $z=0$.}
\label{f6}
\end{figure}
%%%%%%%%%%%%%%%%%%%%%%%%%%%%%%%%%%%%%%%%%%%%%%%%%%%%%%%
%

In Fig.~\ref{f6}, we show $\Delta e_{xc}(z)=e^{RPA}_{xc}(z)-e^{approx}_{xc}(z)$ versus $z/\lambda_F$ 
for a jellium slab with $r_s=2.07$ and several RPA-based approximations for $e^{approx}_{xc}(z)$. The 
RPA-based PBE\cite{YPK} improves considerably the behavior of the RPA-based LDA. The ARPA-GGA 
\cite{CRP11} is a GGA functional that fits the RPA xc energy density of the Airy gas and is 
remarkably accurate for jellium surfaces. The RPA-based GGA++ is the RPA version of the GGA++ of 
Ref.~\onlinecite{CC11}. ($e^{RPA-GGA++}_{xc}=e^{RPA-LSDA}_{xc}F_{xc}(l)$, 
where $l=r_s^2\nabla^2 n/n$ is a reduced Laplacian and $F_{xc}(l)$ is defined in Eq.~(3) of
Ref.~\onlinecite{CC11}.)  Although the GGA++ functional was constructed for the Si crystal, we 
observe that the RPA-based GGA++ improves over the RPA-based LSDA in the bulk near the jellium 
surface showing that it can be a good approximation for systems with small oscillations. (In the 
bulk, close to the jellium surface, there are Friedel oscillations as well as quantum oscillations 
due to the finite thickness of the jellium slab). We note finally that $e^{RPA-L-LSDA}_{xc}$ 
significantly reduces the local error of the RPA-based LSDA near the jellium surface, although by 
construction $E^{RPA-L-LSDA}_{xc}=E^{RPA-LSDA}_{xc}$.

\section{Conclusions}
\label{sec3}

We have presented extensive self-consistent calculations of the {\it exact}-exchange hole and the RPA 
xc hole at a jellium surface.

We have presented a detailed study of the RPA xc hole density at a metal surface. When the 
electron is in the vacuum, its hole remains localized near the surface (its minimum is on 
the image plane) and has damped oscillations in the bulk. We find that the on-top correlation hole is 
accurately described by local and semilocal density-functional approximations, as expected from 
Ref.~\onlinecite{LP} . We also find that for an electron that is localized far outside the surface 
the 
main part of the corresponding xc hole is completely localized at a plane parallel to the surface, 
which is the image plane.

Because of an integration by parts that occurs
in the underlying gradient expansion, a GGA (or meta-GGA) hole is meaningful only after averaging
over the electron density $n(\R)$.\cite{PBW,CPP} This average smooths the sharp cutoffs
used in the construction of the angle-averaged GGA xc hole density. The wavevector analysis of 
the jelium xc surface energy is an important and hard test for the LSDA, GGA, and meta-GGA 
angle-averaged xc hole densities, showing not only the accuracy of the xc hole but also the error 
cancellation between their exchange and correlation contributions. Thus,  Refs.~\onlinecite{PCP} and 
\cite{CPT,CPP} have shown that the 
TPSS meta-GGA~\cite{CPT} and the PBEsol GGA~\cite{CPP} xc hole densities improve 
considerably the accuracy of their LSDA and PBE  counterparts at jellium surfaces, both within RPA
and beyond RPA.\cite{note4}   

The exchange energy density does not have a gradient expansion \cite{PW11}, as does the kinetic 
energy 
density. However the existence of gradient expansion of the xc energy density is still an 
open problem. We use our 
RPA xc hole density to compare the xc energy densities of several approximations. The most accurate 
ones are ARPA GGA of Ref.~\onlinecite{CRP11} and RPA-based L-LSDA of Eq.~ (\ref{e14}).    

\acknowledgments
We thank J. P. Perdew and J. F. Dobson for many valuable discussions and 
suggestions.
J.M.P. acknowledges partial support by the Spanish MEC (grant No. FIS2006-01343 and
CSD2006-53) and the EC 6th framework Network of Excellence NANOQUANTA. 
L.A.C. acknowledges NSF support (Grant No. DMR05-01588).


\begin{thebibliography}{[Vo]}
\bibitem{note1}
$z_0=2\lambda_F$ is sufficiently large for the physical results
to be accurate.
\bibitem{note4}
In Ref. \cite{CPP}, the PBEsol wavevector analysis of a jellium surface had been compared to an 
accurate TDDFT calculation that used the xc kernel of Ref. \cite{PP}.
\end{thebibliography}
\end{document}